\documentclass[aps,prd,preprint,showpacs, showkeys,nofootinbib]{revtex4-1}

\usepackage[T1]{fontenc}
\usepackage[utf8]{inputenc}
\usepackage[english]{babel}
\usepackage{lmodern}

\usepackage{amsmath,amssymb,amsfonts}
\usepackage{graphicx}

\usepackage[colorlinks=true,allcolors=blue]{hyperref}

\numberwithin{equation}{section}

\begin{document}

\title{Quantum Dynamics of a Scalar Particle in Schwarzschild Spacetime using the Generalized Feshbach--Villars Transformation}

\author{Sarra Garah}
\email{sarra.garah@univ-tebessa.dz}
\affiliation{Laboratory of Theoretical and Applied Physics, Echahid Cheikh Larbi Tebessi University, Algeria}

\author{Abdelmalek Boumali}
\email{boumali.abdelmalek@gmail.com}
\affiliation{Laboratory of Theoretical and Applied Physics, Echahid Cheikh Larbi Tebessi University, Algeria}

\date{\today}

\begin{abstract}
In this work, we apply the generalized Feshbach--Villars transformation (GFVT) to spin-0 scalar fields in a Schwarzschild gravitational background. Starting from the covariant Klein--Gordon equation, we reformulate the dynamics in the FV two-component representation, which enables a natural separation of positive- and negative-energy branches. In the far-field approximation, the system exhibits a hydrogen-like bound spectrum, confirming the ability of GFVT to provide a consistent probabilistic interpretation in curved spacetime. We then extend the formalism by introducing a relativistic harmonic oscillator potential, which transforms the radial equation into a biconfluent Heun form. The requirement of square-integrability leads to a discrete oscillator spectrum that remains independent of the gravitational parameter, with gravity appearing only through selection rules on the admissible quantum states. Explicit wave functions, probability densities, and graphical results are presented, illustrating the internal consistency of the method. Overall, this study demonstrates the effectiveness of GFVT as a bridge between relativistic quantum mechanics and curved geometry, and it highlights its potential for future applications in strong gravitational fields.
\end{abstract}

\pacs{03.65.Pm, 04.62.+v, 04.70.-s, 03.65.Ge, 02.30.Gp}
\keywords{Generalized Feshbach--Villars transformation; Feshbach--Villars representation; Klein--Gordon equation; scalar (spin-0) field; Schwarzschild spacetime; biconfluent Heun equation; relativistic harmonic oscillator; positive/negative-energy separation}

\maketitle

\section{Introduction}
The fusion of general relativity and quantum mechanics remains one of the most intriguing and complex challenges in modern theoretical physics \cite{Wald2010,Hayashi1979,Sciama1964,Merzbacher1998,Messiah2014,Greiner1990}. General relativity, formulated by Einstein, describes gravity as the curvature of spacetime, influencing the trajectories of massive objects and the propagation of gravitational waves. In contrast, quantum mechanics, through quantum field theory, explains fundamental interactions at the subatomic scale. However, the attempt to reconcile these two theoretical frameworks into a unified theory of quantum gravity has faced numerous obstacles, particularly when it comes to describing the gravitational effects on relativistic particles, such as those in intense gravitational fields\cite{Thorne1995,HawkingIsrael2010,Choquet2009,Hartle2021}.

In this context, the Feshbach-Villars transformation (GFVT) represents a powerful tool for addressing some of the challenges encountered in the study of relativistic particles. Originally introduced to circumvent the negative energy solutions of the Klein-Gordon equation, this formalism allows for the separation of positive and negative energy components in the solutions, thereby providing a clearer and more coherent interpretation of relativistic quantum states. While GFVT is commonly used in flat spaces or simple geometries, its application in more complex gravitational contexts, such as Schwarzschild backgrounds or black hole fields, remains an area of relatively new exploration\cite{Bouzenada2024,Boumali2024Entropy}.

This study aims to apply the generalized Feshbach-Villars transformation to spin-0 fields in a Schwarzschild gravitational background. The goal is to analyze the effects of spacetime curvature on relativistic particles and investigate how GFVT can enhance our understanding of their quantum dynamics in such environments. Specifically, the study will focus on the impact of gravitational curvature on the energy spectra of spin-0 particles and explore how these results can provide additional insights into the physical properties of extreme gravitational fields, such as those near black holes.

The application of GFVT in this framework represents a significant advancement in understanding relativistic particles in gravitational fields and may potentially open new avenues for integrating quantum gravity into the Standard Model of particle physics.
In this work, we first start from the covariant Klein–Gordon equation in the Schwarzschild geometry and reformulate it within the two-component Feshbach–Villars framework. By adopting Painlevé–Gullstrand coordinates and the tortoise transformation, we derive the effective radial equation that governs the dynamics of spin-0 particles in the far-field regime. We then show that this equation reduces to a Coulomb-like problem, leading to a hydrogen-like bound spectrum. Furthermore, we extend the analysis by introducing a relativistic harmonic oscillator potential into the generalized FV formalism. The resulting differential equation is shown to take the form of a biconfluent Heun equation, whose polynomial truncation yields the discrete oscillator spectrum. Finally, we compute the corresponding wave functions, probability densities, and provide graphical illustrations of the positive- and negative-energy branches, emphasizing the consistency of the GFVT in curved spacetimes and in the presence of external interactions

\section{Klein--Gordon Equation in a Schwarzschild Background}

A real scalar field \(\Phi\) of mass \(m\) satisfies the covariant Klein--Gordon equation with curvature coupling\cite{dietz1976separable}
\begin{equation}\label{eq:KG-cov}
\bigl[\Box + m^2 - \xi\,\mathcal{R}\bigr]\Phi(x)=0,
\end{equation}
where \(\Box=g^{\mu\nu}\nabla_\mu\nabla_\nu\). Outside the central mass we have \(\mathcal{R}=0\). 

We first consider the Schwarzschild metric, written in its standard form as\cite{newman1963empty,dimock1987classical,peters1966perturbations}:
\begin{equation}\label{eq:Schwarzschild-metric}
ds^2=\Bigl(1-\frac{2GM}{r}\Bigr)dt^2-\Bigl(1-\frac{2GM}{r}\Bigr)^{-1}dr^2-r^2\bigl(d\theta^2+\sin^2\theta\,d\phi^2\bigr),
\end{equation}
for which:
\begin{equation}
\sqrt{-g}=r^2\sin\theta,\qquad
g_{\mu\nu}=\mathrm{diag}\Bigl(1-\tfrac{2GM}{r},\,-(1-\tfrac{2GM}{r})^{-1},\,-r^2,\,-r^2\sin^2\theta\Bigr).
\end{equation}
Invoking spherical symmetry and the ansatz\cite{elizalde1988exact}:
\begin{equation}\label{eq:separation}
\Phi(t,r,\theta,\phi)=\frac{1}{r}\,Y_{\ell m}(\theta,\phi)\,e^{-iEt}\,R(r),
\end{equation}
substitution into \eqref{eq:KG-cov} yields the radial equation:
\begin{equation}
\left(1 - \frac{2GM}{r} \right)\frac{d^2R(r)}{dr^2} + \frac{2GM}{r^2}\frac{dR(r)}{dr} + \left[ -\frac{\ell(\ell+1)}{r^2} - m^2 + \frac{E^2}{1 - \dfrac{2GM}{r}} - \frac{2GM}{r^3} \right]R(r) = 0.
\end{equation}

However, the Schwarzschild metric in this form suffers from an apparent coordinate singularity at the event horizon \(r=2GM\). 
Since this singularity is not physical but only due to the choice of coordinates, it is more convenient to work in a regular system. 
For this reason, we shall adopt the Painlevé--Gullstrand (PG) coordinates, which remove the coordinate singularity\cite{volovik2023painleve,herrero2010painleve,leonard2011gravitational,kobayashi2012new}. 
They are obtained through the transformation:
\begin{equation}
dT = dt + \sqrt{\frac{2GM}{r\left(1-2GM/r\right)}}\,dr,
\end{equation}
which leads to the metric:
\begin{equation}
ds^2 = -dT^2 + \left(dr + \sqrt{\frac{2GM}{r}}\, dT \right)^2 + r^2 d\Omega^2,
\end{equation}
or equivalently:
\begin{equation}
ds^2 = \Bigl(1 - \frac{2GM}{r}\Bigr)dT^2 + 2\sqrt{\frac{2GM}{r}}\, dT\,dr - dr^2 - r^2 d\Omega^2.
\end{equation}

In what follows, the Klein--Gordon equation will be reformulated using the Painlevé--Gullstrand representation, which remains regular across the horizon and is therefore more suitable for the analysis of quantum fields in curved spacetime.

\section{Klein--Gordon Equation in Painlevé--Gullstrand Coordinates}

As a preliminary step, let us express the Klein–Gordon equation in Painlevé–Gullstrand coordinates. This choice is particularly convenient since the metric is regular at the horizon:
\begin{equation}
f(r)=1-\frac{2M}{r},\qquad v(r)=\sqrt{\frac{2M}{r}},\qquad
\sqrt{-g}=r^2\sin\theta,
\end{equation}
with the inverse metric components:

\begin{equation}
(g^{\mu\nu})=\begin{pmatrix}
-1 & v & 0 & 0\\
v & f & 0 & 0\\
0 & 0 & \dfrac{1}{r^2} & 0\\
0 & 0 & 0 & \dfrac{1}{r^2\sin^2\theta}
\end{pmatrix}.
\end{equation}

The Klein--Gordon equation $(\Box - m^2)\Phi=0$ reads:
\begin{equation}
\frac{1}{\sqrt{-g}}\partial_\mu\!\Big(\sqrt{-g}\,g^{\mu\nu}\partial_\nu\Phi\Big)-m^2\Phi=0.
\end{equation}

In PG coordinates, using $v'=-\dfrac{v}{2r}$ and $f'=\dfrac{2M}{r^2}$, one obtains:
\begin{equation}\label{eq:KG-PG-PDE}
-\partial_t^2\Phi
+2v\,\partial_{t r}\Phi
+f\,\partial_r^2\Phi
+\frac{2-f}{r}\,\partial_r\Phi
+\frac{3v}{2r}\,\partial_t\Phi
+\frac{1}{r^2}\Delta_{S^2}\Phi
- m^2\Phi=0,
\end{equation}
where :
\begin{equation}
\displaystyle \Delta_{S^2}=\frac{1}{\sin\theta}\partial_\theta\!\big(\sin\theta\,\partial_\theta\big)+\frac{1}{\sin^2\theta}\partial_\phi^2
\end{equation}.

We set:
\begin{equation}
\Phi(t,r,\theta,\phi)=e^{-i\,E t}\,Y_{\ell m}(\theta,\phi)\,\frac{\psi(r)}{r},
\qquad \Delta_{S^2}Y_{\ell m}=-\ell(\ell+1)Y_{\ell m}.
\end{equation}

Substitution into \eqref{eq:KG-PG-PDE} yields the radial ODE (in $r$):
\begin{equation}\label{eq:R-r-PG}
f\,\psi'' + \Big(-2i\,E v+\frac{2M}{r^2}\Big)\psi'
+\Big(\,E^2 - m^2 - \frac{\ell(\ell+1)}{r^2}-\frac{2M}{r^3}
+ \frac{i\,E v}{2r}\Big)\psi=0.
\end{equation}
The imaginary contributions originate from the mixed term $g^{tr}=v$.

We redefine:
\begin{equation}
\psi(r)=e^{\,i\,E S(r)}\,R(r),\qquad {\,S'(r)=\frac{v(r)}{f(r)}\,}.
\end{equation}

Upon substitution, all terms $\propto i\omega$ cancel, and one obtains:
\begin{equation}\label{eq:u-r}
f\,R''+\frac{2M}{r^2}\,R'
+\Big(\frac{\,E^2}{f}-m^2-\frac{\ell(\ell+1)}{r^2}-\frac{2M}{r^3}\Big)R=0.
\end{equation}

Introducing the tortoise coordinate $r_*$\cite{jian2009tortoise}:
\begin{equation}
\frac{dr_*}{dr}=\frac{1}{f(r)},\qquad
\frac{d}{dr_*}=f\,\frac{d}{dr},\qquad
\frac{d^2}{dr_*^2}=f^2\frac{d^2}{dr^2}+ff'\frac{d}{dr},
\end{equation}

and multiplying \eqref{eq:u-r} by $f$, the first-derivative term cancels and one arrives at the Schrödinger-type equation:
\begin{equation}\label{eq:Schro}
{\;\frac{d^2u}{dr_*^2}+\Big[\,E^2 - V_\ell(r)\Big]R=0,\;}
\qquad
{\;V_\ell(r)=f(r)\!\left(m^2+\frac{\ell(\ell+1)}{r^2}+\frac{2M}{r^3}\right).}
\end{equation}

We work in the far region,
\[
\frac{2GM}{r}<1, \qquad r\gg 2GM,
\]
and expand the effective potential, retaining terms up to \(\mathcal{O}(1/r^{2})\) and discarding
\(\mathcal{O}(1/r^{3})\). The effective potential is
\begin{equation}\label{eq:Veff}
V_{\text{eff}}(r)=\Bigl(1-\frac{2GM}{r}\Bigr)\!\left[\frac{l(l+1)}{r^{2}}+\frac{2GM}{r^{3}}+m^{2}\right].
\end{equation}
We separate the massive and centrifugal parts:
\begin{equation}
\Bigl(1-\frac{2GM}{r}\Bigr)\!\left[\frac{l(l+1)}{r^{2}}+m^{2}\right]
+\underbrace{\Bigl(1-\frac{2GM}{r}\Bigr)\frac{2GM}{r^{3}}}_{=\ \mathcal{O}(1/r^{3})\ \text{(discarded)}}.
\end{equation}

Expanding the first factor and keeping up to \(\mathcal{O}(1/r^{2})\) gives
\begin{align}
V_{\text{eff}}(r)
&= \left[\frac{l(l+1)}{r^{2}}+m^{2}\right]
-\frac{2GM}{r}\left[\frac{l(l+1)}{r^{2}}+m^{2}\right]
+\mathcal{O}\!\left(\frac{1}{r^{3}}\right) \nonumber\\
&= m^{2}+\frac{l(l+1)}{r^{2}}-\frac{2GM\,m^{2}}{r}
\ \ +\underbrace{\mathcal{O}\!\left(\frac{1}{r^{3}}\right)}_{\text{includes }-\tfrac{2GM}{r}\!\cdot\!\tfrac{l(l+1)}{r^{2}},\ \tfrac{2GM}{r^{3}}}\,.
\label{eq:Veff_expand}
\end{align}

The radial equation (Regge–Wheeler/Klein–Gordon form) is
\begin{equation}\label{eq:radial_general_rstar}
\frac{d^{2}R}{dr_*^{2}}+\bigl[E^{2}-V_{\text{eff}}(r)\bigr]\,R=0.
\end{equation}
In the far zone we may approximate
\begin{equation}\label{eq:rstar_far}
{\,r_*\simeq r\,}\qquad (r\gg 2GM).
\end{equation}
Substituting \eqref{eq:Veff_expand} into \eqref{eq:radial_general_rstar} and replacing \(r_*\) by \(r\),
we obtain, to the stated order,
\begin{equation}\label{eq:radial_l_general}
\frac{d^{2}R}{dr^{2}}
+\Bigl[E^{2}-m^{2}+\frac{2GM\,m^{2}}{r}-\frac{l(l+1)}{r^{2}}\Bigr]R=0.
\end{equation}

In this subsection we restrict to the s-wave,
\[
{\,l=0\,},
\]
while retaining the \(\bigl(1-\tfrac{2GM}{r}\bigr)\) contribution at order \(1/r\).
The potential becomes:
\begin{equation}\label{eq:Veff_swave}
V_{\text{eff}}^{(l=0)}(r)=m^{2}-\frac{2GM\,m^{2}}{r}
+\mathcal{O}\!\left(\frac{1}{r^{3}}\right),
\end{equation}
and the simplified radial equation reads:
\begin{equation}\label{eq:radial_swave}
{\;
\frac{d^{2}R}{dr^{2}}
+\Bigl[(E^{2}-m^{2})+\frac{2GM\,m^{2}}{r}\Bigr]R=0\; }.
\end{equation}
It is convenient to define:
\[
k^{2}\equiv E^{2}-m^{2},\qquad \alpha\equiv 2GM\,m^{2},
\]
so that \eqref{eq:radial_swave} takes a Coulomb-like form:
\begin{equation}
\frac{d^{2}R}{dr^{2}}+\Bigl[k^{2}+\frac{\alpha}{r}\Bigr]R=0,
\end{equation}

valid for \(r\gg 2GM\) up to \(\mathcal{O}(1/r^{2})\).


Define the Coulomb parameter:
\[
\eta \equiv -\frac{\alpha}{2k}\qquad (k>0\ \text{for }E>m),
\]
and the reduced variable \(z=-2ik r\).
The equation:
\begin{equation}
R''(r)+\Bigl[k^{2}+\frac{\alpha}{r}\Bigr]R(r)=0
\end{equation}

maps to the Whittaker form with parameters \((\kappa,\mu)=(-i\eta,\tfrac12)\).
A basis of solutions is:
\begin{equation}\label{eq:WhittakerSolution}
R(r)=C_{1}\,M_{-i\eta,\,1/2}\!\bigl(-2ik r\bigr)
      +C_{2}\,W_{-i\eta,\,1/2}\!\bigl(-2ik r\bigr).
\end{equation}

\paragraph{Confluent hypergeometric form.}
Using \(M_{\kappa,\mu}(z)=e^{-z/2}z^{\mu+1/2}\,{}_1F_1(\mu-\kappa+\tfrac12,2\mu+1,z)\)
and \(W_{\kappa,\mu}(z)=e^{-z/2}z^{\mu+1/2}\,U(\mu-\kappa+\tfrac12,2\mu+1,z)\),
\eqref{eq:WhittakerSolution} is equivalently:
\begin{equation}\label{eq:KummerSolution}
R(r)=e^{ik r}\,(-2ik r)\Bigl[
A\,{}_1F_1\!\Bigl(1-i\frac{\alpha}{2k},\,2,\,-2ik r\Bigr)
+B\,U\!\Bigl(1-i\frac{\alpha}{2k},\,2,\,-2ik r\Bigr)\Bigr],
\end{equation}

\
Since \(\kappa=\sqrt{m^{2}-E^{2}}\) and using \(\,1-i\frac{\alpha}{2k}=-\,n\), the positive-energy branch only is:
\begin{equation}\label{eq:En_positive}
{\
E_{n}^{(+)} \;=\; m\,\sqrt{\,1-\frac{(GM\,m)^{2}}{(n+1)^{2}}\,}\,,
\qquad n=0,1,2,\dots\ .
}
\end{equation}
This spectrum is real provided \(GM\,m < n+1\).

For \(GM\,m \ll n+1\),
\begin{equation}
E_{n}^{(+)} \;=\; m\left[1-\frac{(GM\,m)^{2}}{2(n+1)^{2}}
+\mathcal{O}\!\left(\frac{(GM\,m)^{4}}{(n+1)^{4}}\right)\right].
\end{equation}

\noindent
This expansion clearly shows that in the weak-gravity regime 
($GMm \ll n+1$), the bound-state energies remain very close to 
the free-particle rest mass $m$, with only a small negative shift. 
The leading correction term, proportional to $(GMm)^2/(n+1)^2$, 
represents a gravitational binding energy analogous to the Coulomb 
correction in the hydrogen atom. 
As the quantum number $n$ increases, this correction becomes 
progressively negligible and the energy tends to $E \simeq m$, 
recovering the free-particle limit. 
This behaviour confirms that the Schwarzschild gravitational field 
induces a hydrogen-like discrete structure, where the lowest states 
are the most affected by the coupling.

\section{Feshbach--Villars Formalism Representation in Flat and Curved Spacetimes}

\subsection{Feshbach-Villars Transformation}
In the Feshbach-Villars representation for spin-0 particles, the goal is to linearize the Klein-Gordon equation (KG), which is a second-order time equation, into a first-order form. This allows a clearer interpretation of positive and negative energies\cite{bouzenada2023statistical,silenko2013scalar}.

The Klein-Gordon equation for a spin-0 particle in Minkowski spacetime, which is of the form \cite{el2003decomposition,bruce1993klein}:

\begin{equation}
\left( \partial_t^2 - \nabla^2 + m^2 \right) \psi(x,t) = 0
\end{equation}

can be transformed using the FV representation. In this representation, the wavefunction \(\psi(x,t)\) is decomposed into two components \(\phi_1(x,t)\) and \(\phi_2(x,t)\), leading to a system of first-order differential equations\cite{bouzenada2024dynamics, Klein1926}:

\begin{equation}
i \frac{\partial \phi_1}{\partial t} = \frac{p^2}{2m} (\phi_1 + \phi_2) + (m + V) \phi_1
\end{equation}

\begin{equation}
i \frac{\partial \phi_2}{\partial t} = - \frac{p^2}{2m} (\phi_1 + \phi_2) - (m - V) \phi_2
\end{equation}

This allows the separation of solutions associated with positive and negative energy. The equation for the total wavefunction is then given by:

\begin{equation}
H_{\text{FV}} \Psi = E \Psi
\end{equation}

with
\begin{equation}
H_{\text{FV}} = \left( \tau_3 + i \tau_2 \right) \frac{p^2}{2m} + m \tau_3 + V(x)
\end{equation}

where \(\tau_3\) and \(\tau_2\) are Pauli matrices, and \(V(x)\) is a potential (such as an electromagnetic potential).

The advantage of this approach is that it allows the separation of positive and negative energies while maintaining a clear probabilistic interpretation of the probability density.

\subsection{Generalization of GFVT}
The Generalized Feshbach-Villars Transformation (GFVT) extends the FV formalism to more complex systems, including curved spacetime and interactions with external fields. The GFVT is useful in general relativity and cosmology, where the spacetime is not flat.
The Klein-Gordon equation in curved spacetime can be written as\cite{bouzenada2023behavior,Gordon1926,Gross1993}:
\begin{equation}
\left( \frac{1}{\sqrt{-g}} \partial_\mu \left( \sqrt{-g} g^{\mu \nu} \partial_\nu \right) + m^2 - \zeta R \right) \Phi(x) = 0
\end{equation}
where \( g_{\mu\nu} \) is the metric tensor, \( R \) is the Ricci scalar, and \( \zeta \) is a coupling constant.
 The generalized Feshbach-Villars transformation (GFVT) is given by:
\begin{equation}
H_{\text{GFVT}} = \tau_z \left( \frac{{N}^2 + T^2}{2N} \right) + i \tau_y \left( \frac{-N^2 + T}{2N} \right) - iY
\end{equation}

where $\mathcal{N}$ is an arbitrary nonzero real parameter, and we have defined $\hat{D} = \frac{\partial}{\partial t} + \mathsf{y}$, with
\begin{equation}
\mathsf{Y} = \frac{1}{2g_{00}} \sqrt{-g} \left\{ \partial_i, \sqrt{-g} g^{0i} \right\}.
\end{equation}

The components of the wave function  in the  GFVT  are provided by:

\begin{equation}
\psi = \phi_{1} + \phi_{2}, \quad i \tilde{D} \psi = N (\phi_{1} - \phi_{2})
\end{equation}

Note that for N = m, the original FV transformations are satisfied:

\begin{equation}
\,T = \frac{1}{g^{00} \sqrt{-g}} \partial_{i} \left( \sqrt{-g} \, g^{ij} \, \partial_{j} \right) + \frac{m^{2} - \zeta R}{g^{00}} - Y^{2}, \quad (i,j = 1,2,3)
\end{equation}
\section{Application ofthe Generalized Feshbach–Villars Transformation (GFVT) in Painlevé–Gullstrand Coordinates}

To avoid the coordinate singularity at the Schwarzschild horizon, 
we adopt the Painlevé–Gullstrand (PG) coordinates. 
This choice ensures a regular description across the horizon and 
facilitates the application of the GFVT framework.
In these coordinates, the line element takes the form:
\begin{equation}
ds^{2} = -f(r)\,dt^{2} + 2\,v(r)\,dt\,dr + dr^{2} + r^{2}d\Omega^{2},
\qquad
f(r)=1-\frac{2GM}{r},\quad v(r)=\sqrt{\frac{2GM}{r}},
\end{equation}

the nonvanishing contravariant components are $g^{00}=-1$, $g^{0r}=v(r)$, $g^{rr}=1$, and $\sqrt{-g}=r^{2}\sin\theta$, yielding:
\begin{equation}
{~
Y\;=\; -\,v(r)\,\partial_{r} \;-\; \frac{3}{4}\,\frac{v(r)}{r},
\qquad
f(r)=1-v^{2}(r).
~}
\label{eq:Y-PG}
\end{equation}

and
\begin{equation}
\Phi =
\begin{pmatrix}\phi_{1}\\[1mm]\phi_{2}\end{pmatrix} e^{-iEt},
\qquad
H_{\text{GFVT}}\,\Phi = i\partial_{t}\Phi,
\end{equation}

one obtains the coupled equations:
\begin{align}
\big(T+N^{2}\big)\phi_{1} + \big(T-N^{2}\big)\phi_{2} &= (2iNY+2NE)\,\phi_{1},\\
-\big(T+N^{2}\big)\phi_{2} - \big(T-N^{2}\big)\phi_{1} &= (2iNY+2NE)\,\phi_{2}.
\end{align}
Eliminating $(\phi_{1}-\phi_{2})$ in favor of $\psi=\phi_{1}+\phi_{2}$ gives the compact spectral form:
\begin{equation}
{~
T\,\psi \;=\; (E+iY)^{2}\,\psi \;=\; \big(E^{2}+2iE\,Y - Y^{2}\big)\psi.
~}
\label{eq:Tpsi}
\end{equation}

Separating variables:
\begin{equation}
\psi(t,r,\theta,\varphi)=\phi(r)\,Y_{\ell m}(\theta,\varphi)\,e^{-i\,E t},
\end{equation}

 the radial equation can be cast as:
\begin{equation}
f(r)\,\phi'' + \left(-2 i E\,v(r) + \frac{2GM}{r^{2}}\right)\phi'
+ \left(E^{2} - m^{2} - \frac{\ell(\ell+1)}{r^{2}} - \frac{2GM}{r^{3}} + \frac{i E\,v(r)}{2r}\right)\phi= 0.
\label{eq:radial-PG}
\end{equation}

To simplify the radial equation , 
we apply a rephasing of the wave function together with the introduction 
of the tortoise coordinate. These steps allow us to rewrite the dynamics 
in a more convenient form, as shown in the following equations.
\cite{jian2009tortoise,li2021long}: 
\begin{equation}
\phi(r)=e^{iE S(r)}\,R(r),\qquad S'(r)=\frac{v(r)}{f(r)},
\quad\text{and}\quad \frac{dr_{*}}{dr}=\frac{1}{f(r)},
\end{equation}
one obtains

\begingroup
\begin{equation}
\frac{d^{2}u}{dr_{*}^{2}} + \Big[E^{2}-V_{\ell}(r)\Big]\,u=0,
V_{\ell}(r)=f(r)\!\left(m^{2}+\frac{\ell(\ell+1)}{r^{2}}+\frac{2GM}{r^{3}}\right).
\end{equation}
\endgroup

Far from the horizon,
\begin{equation}
{~ r_{*}\simeq r, \qquad f(r)\simeq 1, \quad \text{and} \quad \mathcal{O}(1/r^{3})\ \text{negligible}. ~}
\end{equation}

We work in the asymptotic region $r\gg 2GM$ and keep the effect of the redshift factor:
\[
{\,f(r)\equiv 1-\frac{2GM}{r}\,}
\]
to first order in $1/r$, while discarding $\mathcal{O}(1/r^{3})$. 
Specializing immediately to the $s$--wave ($l=0$) yields the effective potential:
\begin{equation}
V_{\rm eff}(r)=f(r)\Big[m^{2}+\underbrace{\frac{2GM}{r^{3}}}_{\mathcal{O}(1/r^{3})}\Big]
= m^{2}-\frac{2GM\,m^{2}}{r}+\mathcal{O}\!\big(1/r^{3}\big).
\label{eq:Veff-s}
\end{equation}

In the far zone one may take $r_{*}\simeq r$ so that radial derivatives with respect to $r_{*}$
can be replaced by derivatives in $r$ at the retained order:
\[
{\,r_{*}\approx r \quad (r\gg 2GM)\,}.
\]

So, we obtain:
\begin{equation}
{\;
\frac{d^{2}R}{dr^{2}}
+\Big[(E^{2}-m^{2})+\frac{2GM\,m^{2}}{r}\Big]\,R(r)=0,
\qquad r\gg 2GM,\; l=0. \;}
\label{eq:far-Coulomb}
\end{equation}
This is Coulomb-like with parameters:
\[
k^{2}\equiv E^{2}-m^{2},\qquad \alpha\equiv 2GM\,m^{2}.
\]

maps to the Whittaker form with parameters \((\kappa,\mu)=(-i\eta,\tfrac12)\).
\cite{whittaker1903partial,temple1956edmund,karel1979functional}A basis of solutions is:
\begin{equation}\label{eq:WhittakerSolution}
R(r)=C_{1}\,M_{-i\eta,\,1/2}\!\bigl(-2ik r\bigr)
      +C_{2}\,W_{-i\eta,\,1/2}\!\bigl(-2ik r\bigr).
\end{equation}

Using \(M_{\kappa,\mu}(z)=e^{-z/2}z^{\mu+1/2}\,{}_1F_1(\mu-\kappa+\tfrac12,2\mu+1,z)\)
and \(W_{\kappa,\mu}(z)=e^{-z/2}z^{\mu+1/2}\,U(\mu-\kappa+\tfrac12,2\mu+1,z)\),
\eqref{eq:WhittakerSolution} is equivalently
\begin{equation}\label{eq:KummerSolution}
R(r)=e^{ik r}\,(-2ik r)\Bigl[
A\,{}_1F_1\!\Bigl(1-i\frac{\alpha}{2k},\,2,\,-2ik r\Bigr)
+B\,U\!\Bigl(1-i\frac{\alpha}{2k},\,2,\,-2ik r\Bigr)\Bigr],
\end{equation}

\
Since \(\kappa=\sqrt{m^{2}-E^{2}}\) and using \(\,1-i\frac{\alpha}{2k}=-\,n\)\cite{gel1992general,seaborn2013hypergeometric,aomoto2011theory}, the energy is:
\begin{equation}\label{eq:En_positive}
{\
E_{n} \;=\pm\; m\,\sqrt{\,1-\frac{(GM\,m)^{2}}{(n+1)^{2}}\,}\,,
\qquad n=0,1,2,\dots\ .
}
\end{equation}
This spectrum is real provided \(GM\,m < n+1\).

For \(GM\,m \ll n+1\),
\[
E_{n}\;=\;\pm m\left[1-\frac{(GM\,m)^{2}}{2(n+1)^{2}}
+\mathcal{O}\!\left(\frac{(GM\,m)^{4}}{(n+1)^{4}}\right)\right].
\]
\begin{equation}
\phi = \frac{1}{2}\left(1 + \frac{E}{m}\right)\psi, \quad \chi = \frac{1}{2}\left(1 - \frac{E}{m}\right)\psi
\end{equation}
Thus:

The spectrum exhibits two symmetric branches: 
a \textbf{positive branch} corresponding to particle states, and a \textbf{negative branch} corresponding to antiparticle states. 
Graphical analysis shows that for small $n$, the energy levels are significantly shifted below the rest mass due to the gravitational interaction, 
indicating the presence of bound states. As $n$ increases, the energy approaches the free-particle limits $\pm m$, 
reflecting the weakening of the gravitational coupling at large distances. 
This structure closely resembles the hydrogen atom spectrum, with gravity playing a role analogous to the Coulomb force. 
Finally, the condition $GM\,m < n+1$ ensures the reality of the spectrum and provides a physical bound for the stability of quantized levels.

\begin{figure}[h!]
  \centering
  \includegraphics[width=0.8\linewidth]{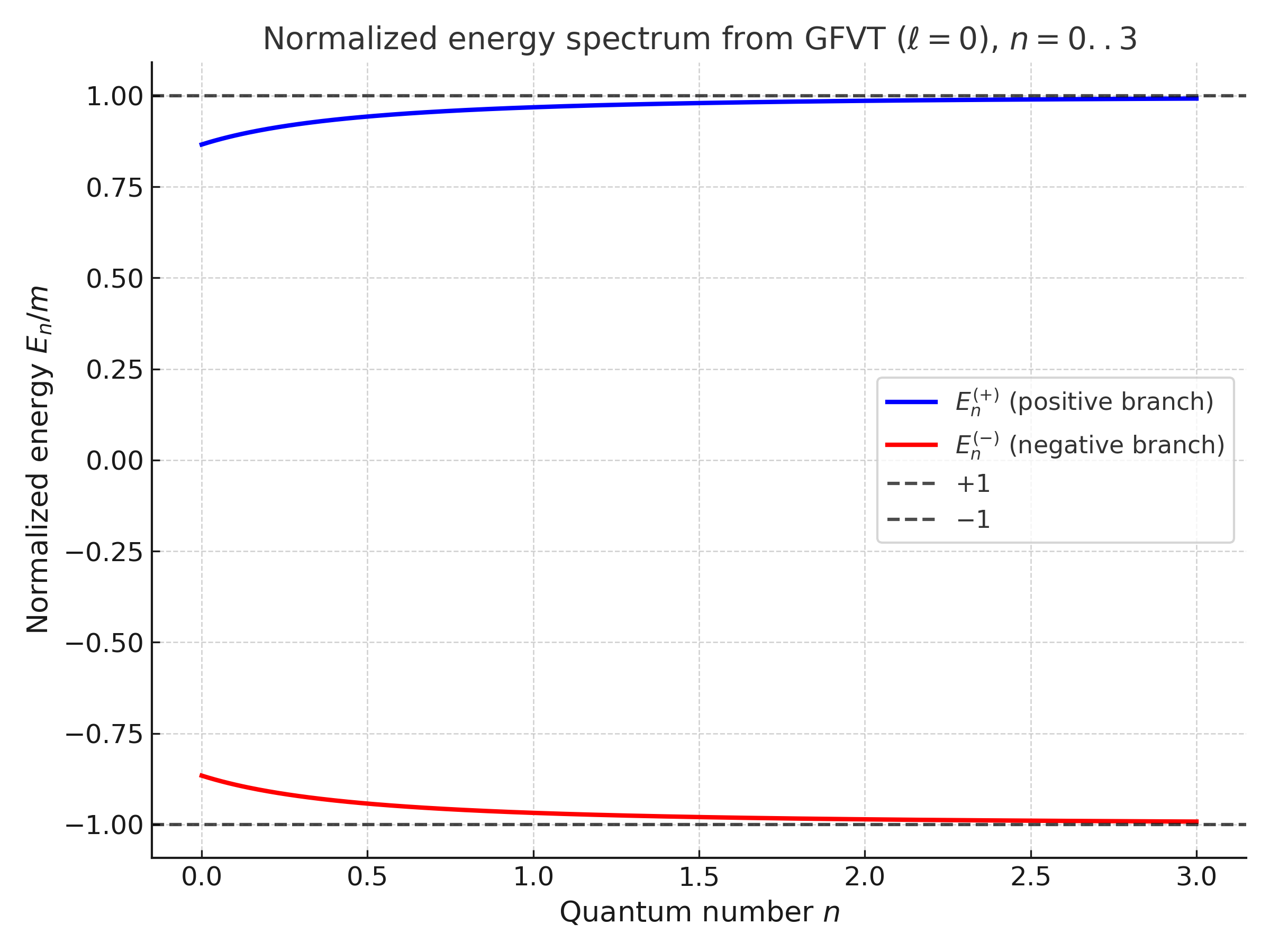}
  \caption{Normalized energy spectrum obtained from the Generalized Feshbach--Villars
  Transformation (GFVT) in the Schwarzschild background for $\ell=0$ and $n \in [0,3]$.
  The blue curve corresponds to the positive-energy branch (particles), while the red
  curve represents the negative-energy branch (antiparticles). Both branches
  asymptotically approach $\pm 1$ (rest-mass energy) as $n$ increases.}
  \label{fig:spectrum}
\end{figure}

\begin{figure}[h!]
  \centering
  \includegraphics[width=0.86\textwidth]{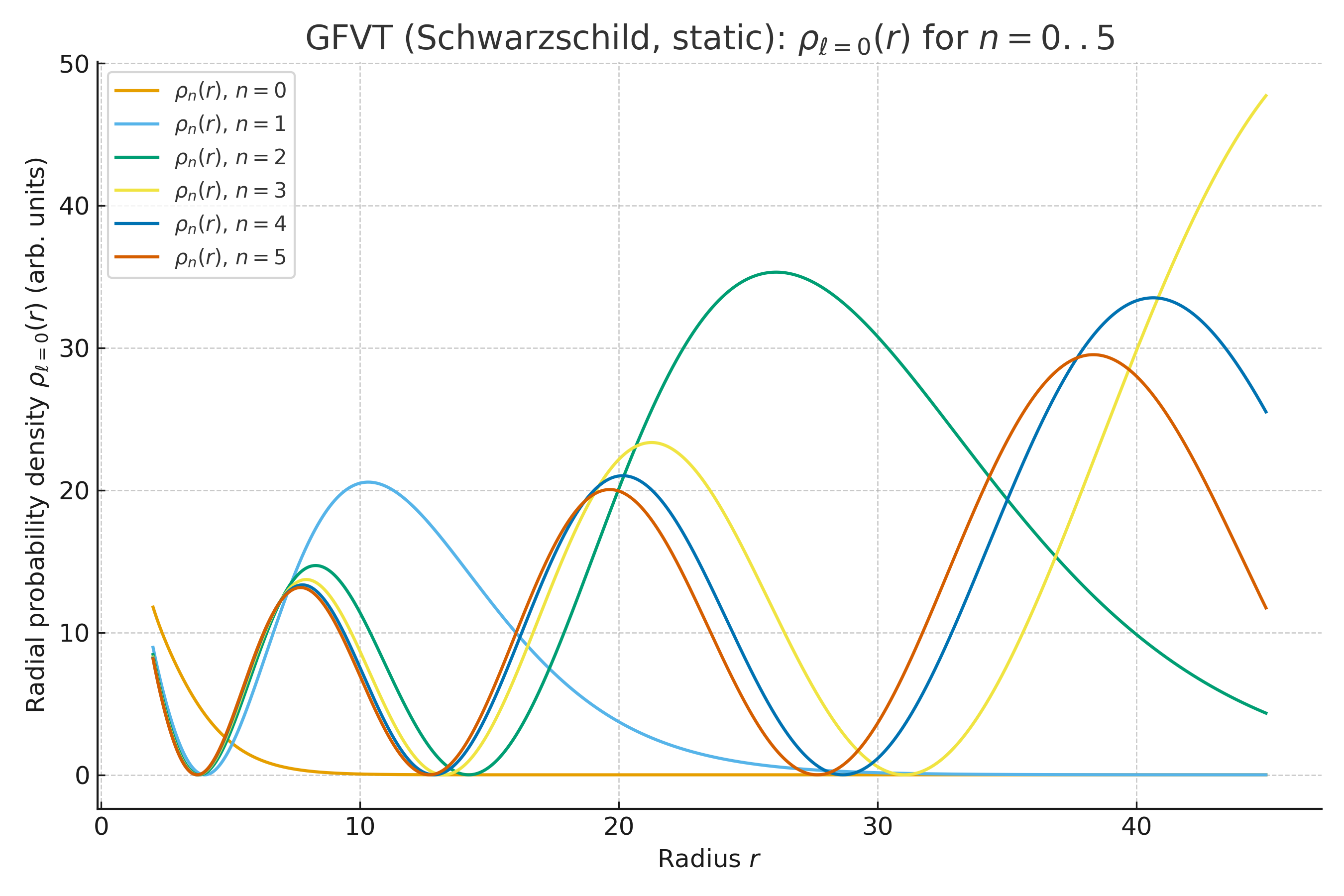}
  \caption{\textbf{Figure 2.} Radial probability density $\rho_{\ell=0}(r)$
  in the Schwarzschild spacetime (static chart) obtained from the GFVT reduction.
  We display the bound-state branches for $n=0,\dots,5$ (parameters used in the plot:
  $m=1$, $GM=0.5$). As $n$ increases, the density extends to larger radii and shows
  the expected sequence of radial nodes. The factor $1/f(r)$ with $f(r)=1-2GM/r$
  encodes the gravitational redshift in the static chart, ensuring a positive density
  for $E>0$ without the coordinate–flux term that appears in PG coordinates.}
  \label{fig:rho-gfvt}
\end{figure}


\subsection*{3. Feshbach--Villars Probability Density}
In the framework of the Feshbach--Villars representation, 
the definition of a proper probability density plays a central role. 
Unlike the standard Klein--Gordon formalism, where the conserved current 
is not positive-definite, the FV approach naturally provides a consistent 
probability interpretation. 
In this context, the positive-definite FV probability density is given by:
\begin{equation}
\rho_{FV} = |\phi|^2 - |\chi|^2
\end{equation}
where the wavefunction components are defined as:
\begin{equation}
\phi = \frac{1}{2}\left(1 + \frac{E}{m}\right)\psi, \quad \chi = \frac{1}{2}\left(1 - \frac{E}{m}\right)\psi
\end{equation}
 the final closed form:
\begin{equation}
{~\rho_n(r)
=4\pi r^2\,\frac{E_n}{m\,f(r)}\;
\bigg|\frac{2\,\kappa_n^{3/2}}{\,n+1\,}\,e^{-\kappa_n r}\,
L_n^{(1)}\!\big(2\kappa_n r\big)\bigg|^2,
\qquad f(r)=1-\frac{2GM}{r},\;
\kappa_n=\frac{GM\,m^2}{n+1}.
~}
\end{equation}

Figure~(2) shows the radial probability densities $\rho_{n}(r)$ for the 
first six bound states with $\ell=0$ in the Schwarzschild background. 
The ground state ($n=0$) is sharply peaked near the origin and decreases 
monotonically with $r$, while the excited states ($n \geq 1$) display an 
increasing number of nodes, as expected for hydrogen-like spectra. 
The maxima of the distributions shift outward with increasing $n$, 
indicating that higher states are spatially more extended. 
This behaviour clearly illustrates the discrete and bound nature of the 
spectrum, as well as the normalizability of the FV wave functions.

\section{Extension to the Harmonic Oscillator Potential}

Within the FV representation of the Klein--Gordon field in the considered metric, and after the standard far-from-horizon approximation, the stationary radial mode \(R(r)\) obeys the second-order ODE:
\begin{equation}
\label{eq:fv-base}
\left[\frac{d^2}{dr^2} + \big(\ E^2 - m^2\big) + \frac{2GM\,m^2}{r}\right] R(r) = 0,
\end{equation}

To add an isotropic harmonic interaction we use the standard FV factorization\cite{boumali2013thermal,boumali2023thermal,boumali2014one}:
\begin{equation}
\Big(\frac{d}{dr} - m\omega r\Big)\Big(\frac{d}{dr} + m\omega r\Big)
= \frac{d^2}{dr^2} + m\omega - m^2\omega^2 r^2,
\end{equation}
which duly accounts for the commutator \([\frac{d}{dr},\, r]=1\). Replacing the free radial operator in \eqref{eq:fv-base} we obtain:
\begin{equation}
\label{eq:radial-osc}
\left[\frac{d^2}{dr^2} - m^2\omega^2 r^2 + \Big(E^2 - m^2 + m\omega + \frac{2GM\,m^2}{r}\Big)\right] R(r) = 0.
\end{equation}

Introduce the dimensionless coordinate \(\zeta=\sqrt{m\omega}\,r\). Equation \eqref{eq:radial-osc} becomes:
\begin{equation}
\label{eq:dimless}
R''(\zeta) + \left(-\zeta^2 + \lambda + \frac{\gamma}{\zeta}\right) R(\zeta) = 0,
\qquad
\lambda=\frac{E^2-m^2+m\omega}{m\omega},\quad
\gamma=\frac{2GM\,m^2}{\sqrt{m\omega}}.
\end{equation}
The large-\(\zeta\) behaviour singles out the Gaussian decay; we therefore factor:
\begin{equation}
\label{eq:ansatz}
R(\zeta)=\zeta\,e^{-\zeta^2/2}\,y(\zeta).
\end{equation}
The reduced function \(y\) satisfies a biconfluent Heun: equation\cite{ishkhanyan2017solutions,karayer2018solution,arriola1991spectral}, so the general solution can be written as
\begin{equation}
\label{eq:heun-gen}
R(\zeta)=\zeta\,e^{-\zeta^2/2}\,HeunB\!\left(1,\,0,\,\lambda,\,-\gamma;\,\zeta\right).
\end{equation}
In general, this contains an \(\mathrm{e}^{+\zeta^2/2}\) component and is not square integrable.

We seek normalizable modes \(R\in L^2(0,\infty)\). Expanding \(y(\zeta)=\sum_{k\ge0} c_k \zeta^k\) leads to the Frobenius recurrence:
\begin{equation}
\label{eq:recurrence}
c_0=1,\quad c_1=-\frac{\gamma}{2},\qquad
(k+2)(k+1)c_{k+2}=(2k+2-\lambda)c_k-\gamma c_{k+1}\quad (k\ge0).
\end{equation}
Physical (decaying) solutions exist \emph{iff} the series truncates at degree \(n\). This yields the two quantization conditions:
\begin{equation}
\label{eq:quant}
{\ \lambda=2n+2\ }\qquad\text{and}\qquad
{\ \Delta_{n+1}(\gamma)=0\ },
\end{equation}
where \(\Delta_{n+1}(\gamma)\) is the tridiagonal determinant obtained by enforcing \(c_{n+1}=0\) in \eqref{eq:recurrence}. The first condition fixes the spectrum,
\begin{equation}
\label{eq:energy}
E_n^2 = m^2 + (2n+1)m\omega,\qquad n=0,1,2,\dots,
\end{equation}
while the second condition is a selection rule involving the gravitational parameter \(\gamma\).

Equation \eqref{eq:energy} shows that the \emph{value} of the energy levels is gravity-independent; gravity contributes only through \(\Delta_{n+1}(\gamma)=0\), selecting which \(n\) are admissible for a given \(\gamma\).

\begin{equation}
\label{eq:fv-components}
\phi = \frac{1}{2}\!\left(1+\frac{E}{m}\right)\psi, \qquad
\chi = \frac{1}{2}\!\left(1-\frac{E}{m}\right)\psi,
\end{equation}
with the two-component spinor \(\Phi=(\phi,\chi)^T\). The FV charge-density observable reads:
\begin{equation}
\label{eq:fv-density}
\rho_{\mathrm{FV}}=\Phi^\dagger \tau_3 \Phi = |\phi|^2 - |\chi|^2.
\end{equation}
Using \(\psi=\phi+\chi\) one readily finds the compact expression:
\begin{equation}
\label{eq:fvo}
\rho_{\mathrm{FVO}}=\frac{|E|}{m}\,|\psi|^2.
\end{equation}
This is the density we use when discussing probability/charge distributions of the oscillator modes.

For illustration we set \(m=\omega=1\) and plot the two energy branches \(E_n=\pm\sqrt{m^2+(2n+1)m\omega}\) as well as the radial densities \(\rho_n(r)\) for \(n=0,\dots,5\).
\begin{figure}[h!]
  \centering
  \includegraphics[width=0.85\textwidth]{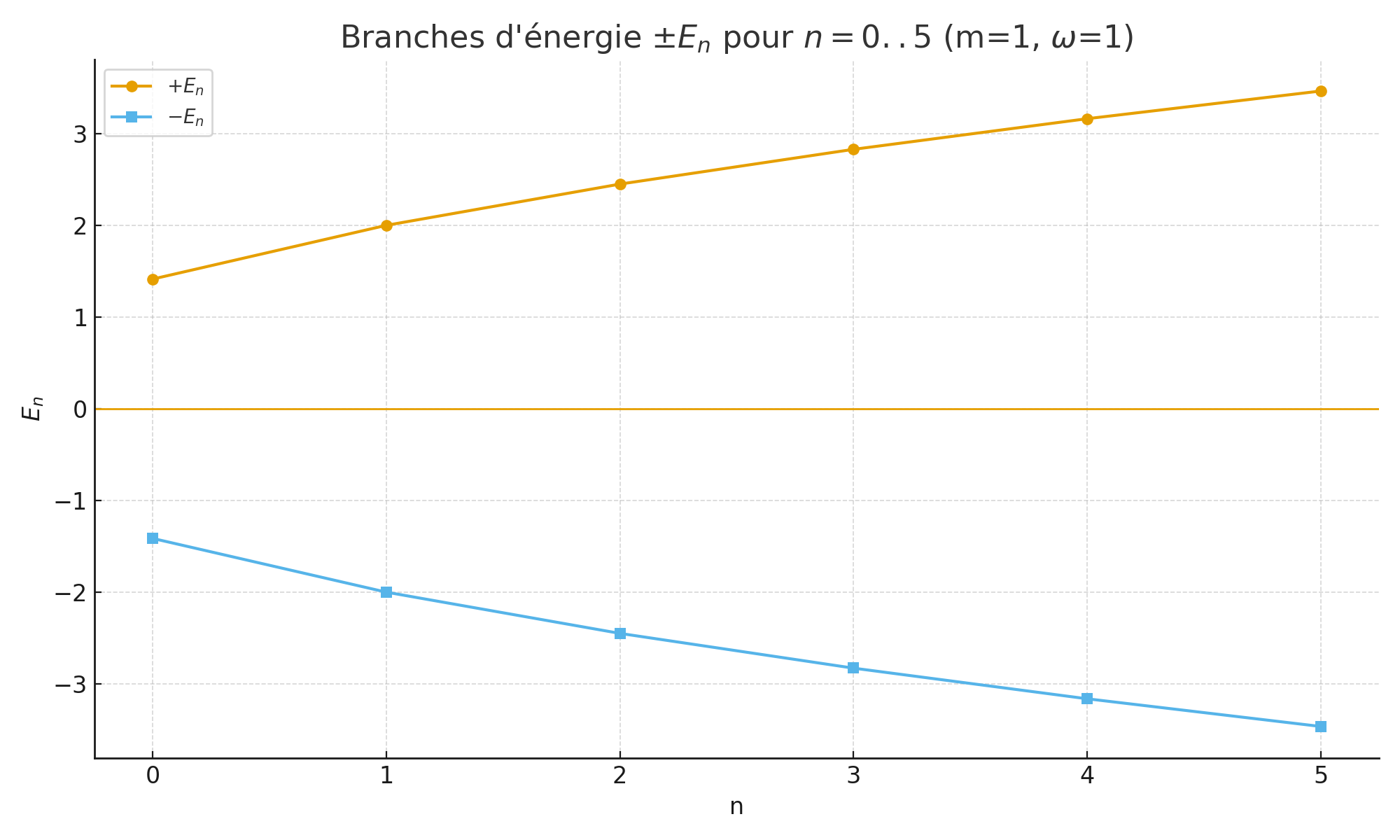}
  \caption{Positive and negative energy branches $ \pm E_n $ for $ n=0,\dots,5 $.}
\end{figure}

\begin{figure}[h!]
  \centering
  \includegraphics[width=0.9\textwidth]{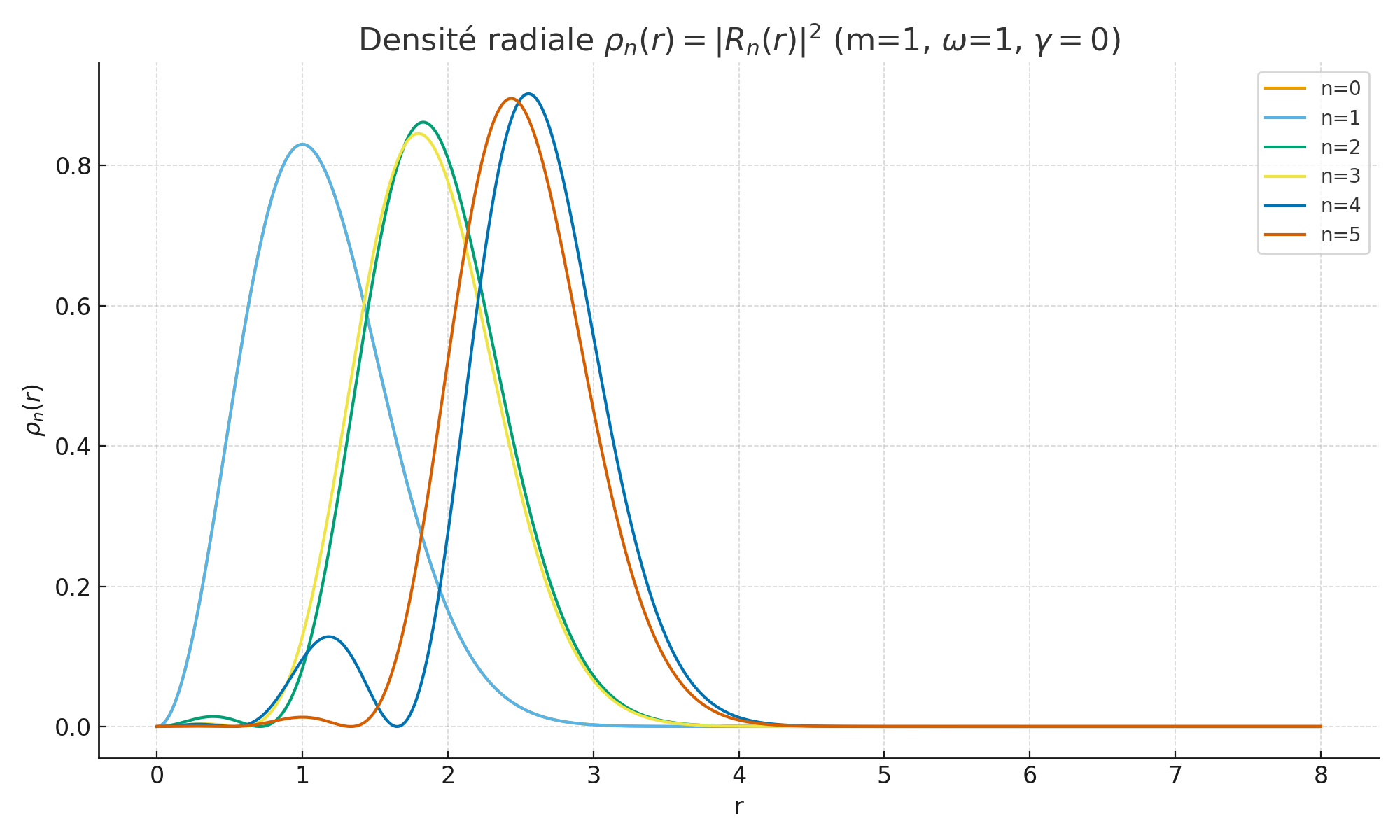}
  \caption{Radial probability densities $ \rho_n(r)=|R_n(r)|^2 $ for $ n=0,\dots,5 $.}
\end{figure}

The harmonic-oscillator sector in the FV formalism reduces the radial problem to a biconfluent Heun equation. Square-integrability imposes polynomial truncation, which yields the gravity-independent spectrum \eqref{eq:energy} with two branches \(E_n=\pm\sqrt{m^2+(2n+1)m\omega}\). Gravity acts only through the selection rule \(\Delta_{n+1}(\gamma)=0\). The FV decomposition \eqref{eq:fv-components} leads to the density observable \eqref{eq:fvo}, which we use to discuss radial distributions.
\subsection{Discussion of the figures}
\label{sec:fig-discussion}

\subsubsection{Energy branches $\pm E_n$}
Figure {3}  displays the two energy branches
\[
E_n=\pm\sqrt{m^2+(2n+1)\,m\omega}, \qquad n=0,1,\dots,5.
\]
\begin{itemize}
  \item \textbf{Symmetry.} The spectrum is symmetric w.r.t.\ zero energy, as expected for the Klein--Gordon/FV framework that carries positive- and negative-energy sectors.
  \item \textbf{Gravity independence.} The \emph{values} of $E_n$ do not depend on the gravitational parameter; gravity only appears as a selection rule through $\Delta_{n+1}(\gamma)=0$, i.e.\ it decides which $n$ are admissible but does not shift the level values.
  \item \textbf{Level spacing.} The spacing $\Delta E_n=E_{n+1}-E_n$ \emph{decreases} with $n$:
  \[
  \Delta E_n=\sqrt{m^2+(2n+3)m\omega}-\sqrt{m^2+(2n+1)m\omega}\sim\frac{m\omega}{\sqrt{m^2+(2n+1)m\omega}},
  \]
  hence a sublinear growth of $E_n$ vs.\ $n$.
  \item \textbf{Nonrelativistic limit.} For $\omega\ll m$,
  \[
  E_n= \pm\Big[m+(n+\tfrac12)\omega - \frac{(n+\tfrac12)^2\omega^2}{2m}+\cdots\Big].
  \]
  After subtracting the rest mass $m$, the $+E_n$ branch reproduces the usual HO ladder $(n+\tfrac12)\omega$ with relativistic corrections $O(\omega^2/m)$.
\end{itemize}

\subsubsection{Radial probability densities $\rho_n(r)=|R_n(r)|^2$}
Figure {4}  shows the normalized radial densities for $n=0,\dots,5$ with $m=\omega=1$ and $\gamma=0$.
\begin{itemize}
  \item \textbf{Behaviour at the origin.} All curves vanish as $r^2$ near $r=0$ because $R_n(r)\propto r$ (for $l=0$), so there is no singularity at the origin.
  \item \textbf{Nodes and outward shift.} The state with quantum number $n$ exhibits exactly $n$ radial nodes. As $n$ increases, the outermost maximum moves outward roughly like $r_{\text{peak}}\propto \sqrt{(2n+1)/(m\omega)}$, reflecting the larger classical turning point.
  \item \textbf{Gaussian tail.} For large $r$,
  \[
  R_n(r)\sim r^{n+1}\,e^{-(m\omega)r^2/2}\quad\Rightarrow\quad \rho_n(r)\sim r^{2n+2}\,e^{-(m\omega)r^2},
  \]
  so the Gaussian decay dominates any polynomial factor and guarantees square integrability.
  \item \textbf{Normalization and orthogonality.} Each curve integrates to unity, $\int_0^\infty |R_n(r)|^2dr=1$, and distinct $n$ are orthogonal with respect to the radial measure $dr$.
  \item \textbf{Effect of the $1/r$ term (qualitative).} The plotted shapes correspond to $\gamma=0$ (``pure'' oscillator). For an \emph{attractive} gravitational term ($\gamma>0$), the density is slightly enhanced at small $r$ and the nodes shift inward; nevertheless, the \emph{energies} stay given by $E_n=\pm\sqrt{m^2+(2n+1)m\omega}$ while $\Delta_{n+1}(\gamma)=0$ may forbid some $n$.
\end{itemize}

\section*{Conclusion}

In this paper, we have applied the generalized Feshbach--Villars transformation (GFVT) to spin-0 scalar fields evolving in a Schwarzschild gravitational background. Starting from the covariant Klein--Gordon equation and reformulating it within the FV two-component framework, we demonstrated that the far-zone radial dynamics can be mapped onto a Coulomb-like problem, yielding a hydrogen-like energy spectrum that exhibits the expected symmetry between positive- and negative-energy branches. This result confirms the ability of the GFVT to provide a clear probabilistic interpretation of scalar dynamics in curved spacetime.

We then extended the analysis by introducing a relativistic harmonic oscillator potential into the FV representation. In this case, the radial equation reduces to a biconfluent Heun form, where the requirement of square-integrability imposes a polynomial truncation, leading to a discrete oscillator spectrum that is independent of the Schwarzschild parameter $GM$. The gravitational field manifests itself only through selection rules restricting the admissible quantum numbers, without altering the oscillator spectrum itself. Explicit wave functions, probability densities, and graphical representations were provided, further illustrating the internal consistency of the GFVT formalism in curved spacetimes.

Overall, this study highlights the usefulness of the generalized FV transformation as a bridge between relativistic quantum mechanics and curved geometry. By demonstrating how the method accommodates both Coulomb-like and oscillator-like interactions in Schwarzschild spacetime, our results open perspectives for applying the GFVT to more general backgrounds and external potentials, thereby contributing to the broader program of exploring quantum dynamics in strong gravitational fields.

\bibliographystyle{unsrt}  
\bibliography{References}

\begin{thebibliography}{10}

\bibitem{Wald2010}
Robert~M. Wald.
\newblock {\em General Relativity}.
\newblock University of Chicago Press, 2010.

\bibitem{Hayashi1979}
Kenji Hayashi and Takeshi Shirafuji.
\newblock New general relativity.
\newblock {\em Physical Review D}, 19(12):3524, 1979.

\bibitem{Sciama1964}
Dennis~W. Sciama.
\newblock The physical structure of general relativity.
\newblock {\em Reviews of Modern Physics}, 36(1):463, 1964.

\bibitem{Merzbacher1998}
Eugen Merzbacher.
\newblock {\em Quantum Mechanics}.
\newblock John Wiley \& Sons, 1998.

\bibitem{Messiah2014}
Albert Messiah.
\newblock {\em Quantum Mechanics}.
\newblock Courier Corporation, 2014.

\bibitem{Greiner1990}
W.~Greiner.
\newblock {\em Relativistic Quantum Mechanics: Wave Equations}.
\newblock Springer, Berlin, 1990.

\bibitem{Thorne1995}
Kip~S. Thorne.
\newblock Gravitational waves.
\newblock {\em arXiv preprint gr-qc/9506086}, 1995.

\bibitem{HawkingIsrael2010}
Stephen Hawking and Werner Israel.
\newblock {\em General Relativity: An Einstein Centenary Survey}.
\newblock Cambridge University Press, 2010.

\bibitem{Choquet2009}
Yvonne Choquet-Bruhat.
\newblock {\em General Relativity and the Einstein Equations}.
\newblock Oxford University Press, 2009.

\bibitem{Hartle2021}
James~B. Hartle.
\newblock {\em Gravity: An Introduction to Einstein’s General Relativity}.
\newblock Cambridge University Press, 2021.

\bibitem{Bouzenada2024}
Abdelmalek Bouzenada, Abdelmalek Boumali, and Faizuddin Ahmed.
\newblock Dynamics of spin-0 (particles–antiparticles) in bonnor–melvin cosmological space-time using the generalized feshbach–villars transformation.
\newblock {\em Nuclear Physics B}, 1007:116682, 2024.

\bibitem{Boumali2024Entropy}
A.~Boumali, A.~Hamla, and Y.~Chargui.
\newblock Determination of shannon entropy and fisher information of the feshbach--villars oscillator for spin-0 particles.
\newblock {\em International Journal of Theoretical Physics}, 63(8):200, 2024.

\bibitem{dietz1976separable}
W~Dietz.
\newblock Separable coordinate systems for the hamilton-jacobi, klein-gordon and wave equations in curved spaces.
\newblock {\em Journal of Physics A: Mathematical and General}, 9(4):519, 1976.

\bibitem{newman1963empty}
Ezra Newman, L~Tamburino, and Theodore Unti.
\newblock Empty-space generalization of the schwarzschild metric.
\newblock {\em Journal of Mathematical Physics}, 4(7):915--923, 1963.

\bibitem{dimock1987classical}
Jonathan Dimock and Bernard~S Kay.
\newblock Classical and quantum scattering theory for linear scalar fields on the schwarzschild metric i.
\newblock {\em Annals of physics}, 175(2):366--426, 1987.

\bibitem{peters1966perturbations}
PC~Peters.
\newblock Perturbations in the schwarzschild metric.
\newblock {\em Physical Review}, 146(4):938, 1966.

\bibitem{elizalde1988exact}
Emili Elizalde.
\newblock Exact solutions of the massive klein-gordon-schwarzschild equation.
\newblock {\em Physical Review D}, 37(8):2127, 1988.

\bibitem{volovik2023painleve}
GE~Volovik.
\newblock Painlev{\'e}--gullstrand coordinates for schwarzschild--de sitter spacetime.
\newblock {\em Annals of Physics}, 449:169219, 2023.

\bibitem{herrero2010painleve}
Alicia Herrero and Juan~Antonio Morales-Lladosa.
\newblock Painlev{\'e}--gullstrand synchronizations in spherical symmetry.
\newblock {\em Classical and Quantum Gravity}, 27(17):175007, 2010.

\bibitem{leonard2011gravitational}
C~Danielle Leonard, Jonathan Ziprick, Gabor Kunstatter, and Robert~B Mann.
\newblock Gravitational collapse of k-essence matter in painlev{\'e}-gullstrand coordinates.
\newblock {\em Journal of High Energy Physics}, 2011(10):1--21, 2011.

\bibitem{kobayashi2012new}
Tsutomu Kobayashi, Masaru Siino, Masahide Yamaguchi, and Daisuke Yoshida.
\newblock New cosmological solutions in massive gravity.
\newblock {\em Physical Review D—Particles, Fields, Gravitation, and Cosmology}, 86(6):061505, 2012.

\bibitem{jian2009tortoise}
Yang Jian, Zhao Zheng, Tian Gui-Hua, and Liu Wen-Biao.
\newblock Tortoise coordinates and hawking radiation in a dynamical spherically symmetric spacetime.
\newblock {\em Chinese Physics Letters}, 26(12):120401, 2009.

\bibitem{bouzenada2023statistical}
Abdelmalek Bouzenada and Abdelmalek Boumali.
\newblock Statistical properties of the two dimensional feshbach--villars oscillator (fvo) in the rotating cosmic string space--time.
\newblock {\em Annals of Physics}, 452:169302, 2023.

\bibitem{silenko2013scalar}
Alexander~J Silenko.
\newblock Scalar particle in general inertial and gravitational fields<? format?> and conformal invariance revisited.
\newblock {\em Physical Review D—Particles, Fields, Gravitation, and Cosmology}, 88(4):045004, 2013.

\bibitem{el2003decomposition}
Salah~M El-Sayed.
\newblock The decomposition method for studying the klein--gordon equation.
\newblock {\em Chaos, Solitons \& Fractals}, 18(5):1025--1030, 2003.

\bibitem{bruce1993klein}
S~Bruce and P~Minning.
\newblock The klein-gordon oscillator.
\newblock {\em Il Nuovo Cimento A (1965-1970)}, 106(5):711--713, 1993.

\bibitem{bouzenada2024dynamics}
Abdelmalek Bouzenada, Abdelmalek Boumali, and Faizuddin Ahmed.
\newblock Dynamics of spin-0 (particles-antiparticles) in bonnor-melvin cosmological space-time using the generalized feshbach-villars transformation.
\newblock {\em Nuclear Physics B}, 1007:116682, 2024.

\bibitem{Klein1926}
O.~Klein.
\newblock Quantum theory and five-dimensional theory of relativity.
\newblock {\em Zeitschrift f{\"u}r Physik}, 37:895, 1926.

\bibitem{bouzenada2023behavior}
Abdelmalek Bouzenada, Abdelmalek Boumali, Omar Mustafa, and Hassan Hassanabadi.
\newblock Behavior of the feshbach-villars oscillator (fvo) in g$\backslash$" urses space-time under coulomb-type potential.
\newblock {\em arXiv preprint arXiv:2304.12496}, 2023.

\bibitem{Gordon1926}
W.~Gordon.
\newblock Der comptoneffekt nach der schr{\"o}dingerschen theorie.
\newblock {\em Zeitschrift f{\"u}r Physik}, 40:117, 1926.

\bibitem{Gross1993}
F.~L. Gross.
\newblock {\em Relativistic Quantum Mechanics and Field Theory}.
\newblock Wiley, 1993.

\bibitem{li2021long}
Wen-Du Li and Wu-Sheng Dai.
\newblock Long-range potential scattering: Converting long-range potential to short-range potential by tortoise coordinate.
\newblock {\em Journal of Mathematical Physics}, 62(12), 2021.

\bibitem{whittaker1903partial}
Edmund~T Whittaker.
\newblock On the partial differential equations of mathematical physics.
\newblock {\em Mathematische Annalen}, 57(3):333--355, 1903.

\bibitem{temple1956edmund}
George Frederick~James Temple.
\newblock Edmund taylor whittaker, 1873-1956, 1956.

\bibitem{karel1979functional}
Martin~L Karel.
\newblock Functional equations of whittaker functions on p-adic groups.
\newblock {\em American Journal of Mathematics}, 101(6):1303--1325, 1979.

\bibitem{gel1992general}
Izrail~Moiseevich Gel'fand, Mark~Iosifovich Graev, and Vladimir~Solomonovich Retakh.
\newblock General hypergeometric systems of equations and series of hypergeometric type.
\newblock {\em Russian Mathematical Surveys}, 47(4):1, 1992.

\bibitem{seaborn2013hypergeometric}
James~B Seaborn.
\newblock {\em Hypergeometric functions and their applications}, volume~8.
\newblock Springer Science \& Business Media, 2013.

\bibitem{aomoto2011theory}
Kazuhiko Aomoto.
\newblock {\em Theory of hypergeometric functions}.
\newblock Springer, 2011.

\bibitem{boumali2013thermal}
Abdelmalek Boumali and Hassan Hassanabadi.
\newblock The thermal properties of a two-dimensional dirac oscillator under an external magnetic field.
\newblock {\em The European Physical Journal Plus}, 128(10):124, 2013.

\bibitem{boumali2023thermal}
Abdelmalek Boumali and Nabil Korichi.
\newblock On the thermal properties of the one-dimensional space fractional duffin--kemmer--petiau oscillator.
\newblock {\em Physics of Particles and Nuclei Letters}, 20(2):100--111, 2023.

\bibitem{boumali2014one}
Abdelmalek Boumali.
\newblock The one-dimensional thermal properties for the relativistic harmonic oscillators.
\newblock {\em arXiv preprint arXiv:1409.6205}, 2014.

\bibitem{ishkhanyan2017solutions}
TA~Ishkhanyan and AM~Ishkhanyan.
\newblock Solutions of the bi-confluent heun equation in terms of the hermite functions.
\newblock {\em Annals of Physics}, 383:79--91, 2017.

\bibitem{karayer2018solution}
H~Karayer, D~Demirhan, and F~B{\"u}y{\"u}kk{\i}l{\i}{\c{c}}.
\newblock Solution of schr{\"o}dinger equation for two different potentials using extended nikiforov-uvarov method and polynomial solutions of biconfluent heun equation.
\newblock {\em Journal of Mathematical Physics}, 59(5), 2018.

\bibitem{arriola1991spectral}
E~Ruiz Arriola, A~Zarzo, and JS1136921 Dehesa.
\newblock Spectral properties of the biconfluent heun differential equation.
\newblock {\em Journal of computational and applied mathematics}, 37(1-3):161--169, 1991.

\end{thebibliography}

\end{document}